# Green IT as a tool for design cloud-oriented sustainable learning environment of a higher education institution

*Tetiana* Vakaliuk[1,*], *Dmitry* Antoniuk[1], *Andrii* Morozov[1], *Mariia* Medvedieva[2], and *Mykhailo* Medvediev[3]

[1]Zhytomyr Polytechnic State University, Department of Software Engineering, Zhytomyr, 10005, Ukraine
[2]Pavlo Tychyna Uman State Pedagogical University, Department of informatics and information and communication technologies, Uman, 20300, Ukraine
[3]ADA University, School of Information Technologies and Engineering, Baku, AZ1008, Azerbaijan

**Abstract.** The paper proposes the use of green IT as a tool for designing a cloud-oriented sustainable learning environment for a higher education institution. The article substantiates the expediency of designing such an environment as a prerequisite for the sustainable development of Ukraine. It is established that one of the goals of Ukraine's sustainable development for 2030 is to provide fair quality education and to promote lifelong learning opportunities for all. Green IT is a set of approaches related to sustainable computing and information technology. The work of foreign scientists was analyzed, which considered the issues of designing the learning environment using green computing. As a result, Cloud LMS has been established that cloud LMS is a type of green IT and can serve as a tool for designing a cloud-oriented sustainable learning environment of a higher education institution. A model of a cloud-oriented sustainable learning environment of a higher education institution using cloud LMS is proposed. The application of a cloud-oriented sustainable learning environment will provide such capabilities: keep electronic journals; use on-line services; conduct correspondence, assessment of knowledge on-line; and more. And all of the above is the key to a sustainable development of the learning environment.

## 1 Introduction

Sustainable human development is a process of change in which the exploitation of resources, the direction of investments, the orientation of technological development of mankind and institutional changes agreed with each other that strengthen the current and future potential to meet human needs and aspirations [1, p. 7].

All these years, there has been ongoing work aimed at bringing to the world public an understanding of the nature and magnitude of the danger, coordinating the activities of international organizations, governments, non-governmental organizations, scientists in the search and implementation of effective measures to reduce the level of global threats to humanity.

To create conditions for sustainable development for the country, it is necessary to focus on the implementation of such important components as:
- economical – represents the economical use of all types of resources aimed at reducing pressure on natural ecosystems;
- ecological – a series of actions that lead to the restoration of the original state of the environment to a level that does not harm either human health or natural ecosystems and helps to maximize improvement;
- social – an integral part, which provides for the improvement of the quality of life of a person, because the person acts as the main cause of changes occurring in society.

According to the 2030 Sustainable Development Strategy of Ukraine, one of the goals of sustainable development is to provide equitable quality education and to promote lifelong learning opportunities for all, which envisages ensuring lifelong access to quality vocational and higher education for the whole population, significantly increase the number of young and adult people who have the socially necessary skills, including vocational and technical skills for employment, decent work, and entrepreneurial employment what activity [2].

At the same time, green information technology is a "set of approaches related to environmentally friendly computing and information technology. It is the science and practice of designing, manufacturing, using and utilizing computers, servers and their subsystems, such as monitors, printers, devices storage of data, networks and communication systems – effectively and with minimal or zero environmental impact" [3].

## 2 Literature review

Sustainable computing in recent years increasingly involved scientists from around the world [4-17]. In particular, V. Kharchenko and O. Illiashenko offers concepts of green IT engineering: taxonomy, principles, and implementation [6], Y. Kondratenko, O. Korobko,

---

[*] Corresponding author: tetianavakaliuk@gmail.com

O. Kozlov views PLC-oriented systems for data acquisition and supervisory control of environment-friendly energy-saving technologies [7], N. Boroujerdi and S. Nazem considering cloud computing as changing cogitation about computing [8], V. Kharchenko, Y. Kondratenko, J. Kacprzyk offers concepts, models, complex systems architectures, studies in systems, decision, and control of Green IT engineering [9], Y. Kondratenko, V. Korobko, O. Korobko, G. Kondratenko, O. Kozlov considering Green IT approach to design and optimization of thermoacoustic waste heat utilization plant-based on soft computing [10].

Other scientists – J. Drozd, A. Drozd, S. Antoshchuk view resource-oriented approach of Green IT engineering [11], V. Hahanov, E. Litvinova, S. Chumachenko offers green cyber-physical computing as sustainable development model [12], N. Bardis view secure, green implementation of modular arithmetic operations for IoT and cloud applications [13], and N. Doukas considering technologies for Greener Internet of Things systems [14].

K. Palanivel and S. Kuppuswami in their study proposed a cloud-oriented green computing architecture for e-learning applications: COGALA [15]. This is due to the fact that the rapid development of cloud technologies in the future implies a lack of high-speed cloud-oriented architectures for educational institutions [15]. They also offer their models of cloud-oriented e-learning architecture (see Figure 1) and cloud-oriented green computing architecture for e-learning (see Figure 2) [15].

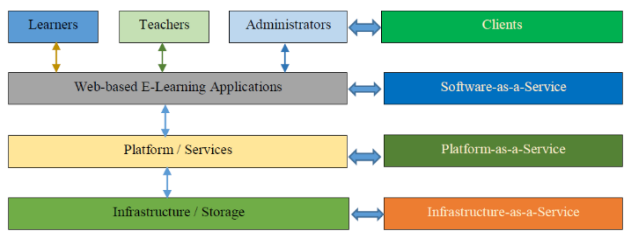

**Fig. 1.** Cloud-oriented e-learning architecture model (K. Palaniwell and S. Kupuswami).

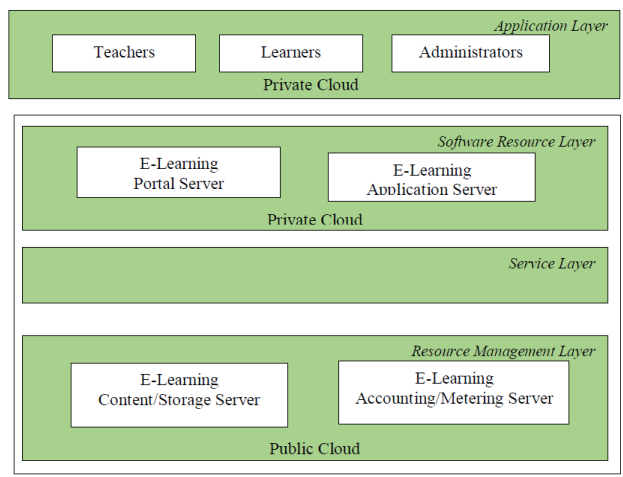

**Fig. 2.** Cloud-oriented green computing architecture model for e-learning (K. Palanivel and S. Kupuswami).

Figure 1 presents the architecture of an e-learning system based on a cloud-oriented architecture, according to which the model is divided into three levels, including infrastructure, platforms, and applications. At the infrastructure level, training resources from the traditional system are transferred to the cloud database instead of the usual database. The platform level involves the use of a new e-learning system based on CMS (Content Management System) and LMS (Learning Management System). These components are important in the model because they are designed to mediate between a cloud-oriented database and native applications. Finally, the application layer was designed to interact with the client (student, teacher, administrator) [15].

K. Palanivel and S. Kuppuswami also propose a model of a cloud-oriented green computing architecture for e-learning (see Fig. 2) [15]. It is divided into four levels: the public cloud resource management layer and containing the e-learning content/storage server on the e-learning accounting/metering server; level of service; the level of software resources hosted in a private cloud consisting of a portal server and an e-learning application server; application level, which is also hosted in a private cloud (at this level, users are teachers, students, administrators) [15].

## 3 Results

However, the problem of creating a cloud-oriented sustainable education environment for higher education institutions is nowadays needed to be addressed. That is why the structure and functioning of the cloud-oriented sustainable learning environment higher education institution that underlies the design of such an environment remain unexplored.

According to the V. Yu. Bykov and V. H. Kremen, to design the learning environment means to theoretically investigate the essential target and content-technological (methodical) aspects of the educational process which should be carried out in the learning environment, and on this basis to describe the necessary for this purpose the structure and structure of the learning environment (its statics and dynamics, including to predict and take into account the development of the structure of the learning environment, the influence and peculiarities of the relationship of the learning environment components with other elements of the environment, with the elements of the environment) in accordance with the dynamics of the development of its goals creation and use, and limitations of psychological and educational, scientific and resource nature [16, p. 7].

Summarizing the above interpretation, it can be argued that the theoretical study of the learning environment is to create a model that will provide an idea of the learning environment in which the cooperation and communication of all participants in the educational process.

Specialized platforms, such as LMS, are constantly being created at the higher education institution to address the challenge of deploying higher education institution education and training systems online and in designing a cloud-oriented learning environment. They are used to develop, manage, and distribute educational online sharing materials. The materials are placed in a learning environment with the definition of the sequence of their study. LMS is comprised of a variety of individual tasks, small-group projects, and learning elements for all students, both content-oriented and communicative.

There are several training management systems that allow learning using the Internet. Thus, the learning process can be done in real-time by organizing online lectures and seminars.

LMS in the form of use is conventionally divided into two types [17, p. 117]:

1. **LMS as software** that is intended for installation on your own higher education institution servers. The use of LMS of this type implies receiving an appropriate service from an IaaS cloud provider. The operation of such LMS requires the availability of appropriate personnel and software.

2. **LMS as a cloud platform** created by a provider used by users to manage the educational process. The use of LMS of this type involves obtaining an appropriate service from an higher education institution using a SaaS cloud service model. Due to this, all the basic functions of maintenance and technical support are assigned to a specific provider.

As a result of our research, consider that cloud LMS is a type of Green IT.

After analyzing the main scientific papers on the subject, a generalization and model of a cloud-oriented sustainable learning environment of a higher education institution in the following form were summarized (see Fig. 3).

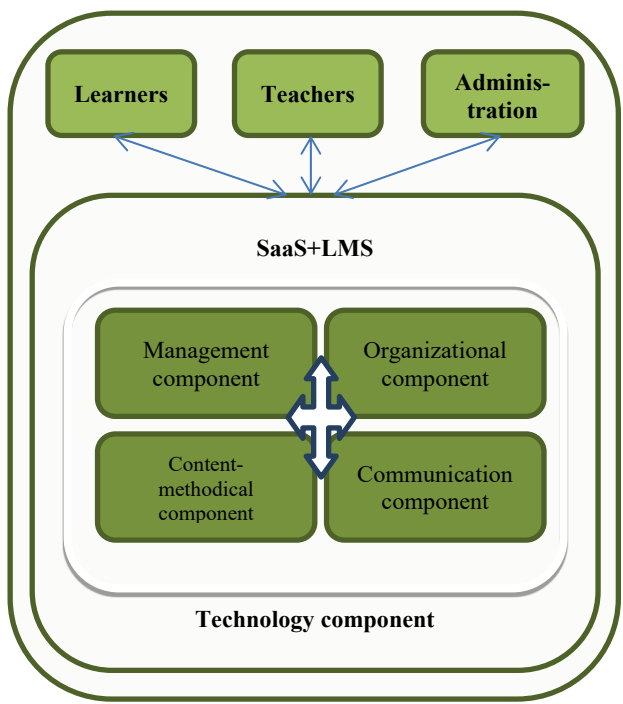

**Fig. 3.** Model of a cloud-oriented sustainable learning environment of a higher education institution

Since learning environment "is an artificially constructed system, structure, and components which create the conditions necessary to achieve the objectives of the educational process ... and learning environment structure determines its internal organization, relationship and interdependence between its elements" [18, p. 3], then the proposed model (see Fig. 3) is oriented towards the achievement of the learning objectives, which are reflected in the industry standards of higher education through all structural constituents of the cloud-oriented sustainable learning environment of the higher education institution.

To achieve the learning objectives, a cloud-oriented sustainable learning environment in a higher education institution should fulfill the following functions:

- *management* – management of the educational process of preparation of future professionals;
- *organizational* – organizing the actual learning process through the distribution of access rights and the distribution of communities of entities;
- *educational* – submission of educational materials, as well as practical and laboratory work;
- *advisory* – providing on-line consultations to students and groups of students;
- *communication* – the presence of subject-subject interaction, as well as the possibility of communication between the subjects, which is also the implementation of a feedback mechanism;
- *controlling* – the availability of an electronic journal and the ability to evaluate works online;
- *developmental* – development of students' information and communication competencies;
- *systematic* – systematization of materials.

These functions are possible in the presence of the following structural components of a cloud-oriented sustainable learning environment of a higher education institution.

***The technological component*** is implemented through the use of a cloud-oriented learning management system and combines managerial, organizational, content and methodical and communication components that are interconnected with each other.

A cloud-oriented learning management system will be understood to mean a system that facilitates group collaboration between faculty and students, the development, management, and dissemination of educational materials with the sharing of the educational process through cloud technology.

Consider in detail each of its components.

The management component provides for the use of cloud-oriented learning management tools and cloud-oriented learning performance assessment tools in the educational process of a higher education institution. Also within this component should be defined as the disciplines provided by the curriculum of the relevant specialty.

The cloud-oriented sustainable learning environment of the institution of higher education provides great opportunities for control of students' educational activities: laboratory work, testing, independent work, joint projects, control works, etc.

***Organizational component.***

The successful functioning of a cloud-oriented sustainable learning environment in a higher education institution is possible provided that a logical distribution of user access rights is made. The following users are highlighted: admin; lecturer (including representatives of higher education institution administration); students.

Note that each user group has limited access rights to its own cloud-oriented sustainable learning environment of a higher education institution. Students may have the opportunity to read or edit a variety of teaching and teaching materials.

Each teacher in the cloud-oriented sustainable learning environment of the higher education institution is provided with his own cloud-oriented office, where he can store all the necessary materials for successful and high-quality classes: educational and work programs of disciplines, textbooks, manuals, lectures, laboratory materials, exam requirements, coursework and diploma instruction, etc.

With such advanced functionality, teachers can provide access to all the necessary materials, allowing students to complete: joint projects, independent work, research work, etc.

The following types of documents should be developed, structured and made available to faculty members in a cloud-oriented sustainable learning environment for higher education institutions:

− regulatory documents both at the national level and at the university level;
− curricula for the preparation of future professionals;
− methodical recommendations on filling of educational-methodical complexes of disciplines.

When designing this component, it is necessary to take into account all the necessary components: the schedule of classes, terms of delivery of work, the plan of work of the department, faculty, announcements, discussions, contacts, etc. And given the specifics of cloud technologies, it is advisable that a cloud-oriented sustainable learning environment in a higher education institution also has internal communication tools (a kind of its social network), a forum (to engage all subjects in the educational process). Also positive is the creation of photo albums of groups, departments, which would contain materials of all events taking place in higher education institution.

In order to control the success and quality activities of teachers in the educational process in the cloud-oriented sustainable learning environment of a higher education institution, an administration should be present. The administration controls: the conformity of the available subjects in the cloud-oriented sustainable learning environment of a higher education institution to the curricula for the training of specialists in the relevant specialty, the conformity of the educational materials placed in such environment to the curricula of all disciplines.

This provides the opportunity: keep electronic journals; testing, and assessment of knowledge on-line; and more.

***Content-methodical component.***

Learning objectives affect the content of the training, which in combination influences the choice of tools, methods, and forms of organization of education in higher education.

The content of a cloud-oriented sustainable learning environment of a higher education institution is in line with all concepts of education, industry standards of higher education, curricula for the training of relevant specialists, educational and work programs of the disciplines envisaged for mastering.

In this case, each component of the methodological system is divided into traditional and cloud-oriented components. It should be noted that cloud-oriented learning tools complement the traditional higher education educational process.

The use of cloud-oriented learning tools facilitates the identification of traditional and cloud-oriented methods and forms of the learning organization. By cloud-oriented methods and forms of training organization, we will understand such methods and forms that are implemented in the educational process with the use of cloud ICT infrastructure.

Proper selection of modern methods, forms and tools of teaching (cloud-oriented and traditional) in accordance with the goals of a certain discipline promotes the development of cognitive abilities of students, the development of creative and logical thinking, the formation of skills to use the acquired knowledge in practice, the formation of necessary professional competences, including information and communication competence as a component of professional competence of future specialists for further creative activity.

This provides the opportunity: use on-line services for the educational process; possibility of distance learning, library of books, textbooks, media files; file repositories and more.

***Communication component.***

Communicating between subjects of training activities implemented directly with each other and through cloud-oriented communications in a cloud-oriented sustainable learning environment of a higher education institution.

Important in the communication component is the selection of modes of communication. According to V. Yu. Bykov, they are synchronous and asynchronous [19, p. 323]. The synchronous mode of communication implies simultaneous interaction of the subjects of learning at one time, and the asynchronous one assumes independence of the interaction time of the subjects of the educational process.

The successful and quality functioning cloud-oriented sustainable learning environment of a higher education institution should be provided both modes of communication.

Considering the interaction of the participants in the learning process in a cloud-oriented sustainable learning environment of a higher education institution, it is first of all necessary to determine the subjects of interaction. In our case, the subjects of interaction are the student, the teacher, and the scientific supervisor.

It should be noted at the outset that the scientific advisor is distinguished by a separate entity because the curricula of higher education institutions include such types of work as the writing of coursework and diploma projects (papers), in which the scientific supervisor plays a leading role.

The subjects of interaction define the links of interaction in the cloud-oriented sustainable learning environment of a higher education institution, which should include: student-student, student-teacher, teacher-student-student, student-scientific supervisor. It should be noted that the interaction of teacher-student-student is one of the defining elements in the educational process of a higher education institution. After all, it depends on her interpersonal relationships not only students but also students with the teacher.

Different levels of interaction covering the various types and forms of interaction. In particular, the following types of interaction are highlighted: individual activity, group interaction, subgroup interaction, and couples interaction.

In a cloud-oriented sustainable learning environment of a higher education institution, students independently perform tasks (individual work), perform joint projects, discuss problems (interaction in pairs, subgroups), use the process of learning (interaction in pairs, subgroups, groups), communicate with each other (interaction in groups).

The main forms of interaction between the subjects of the educational process in such an environment include: information, consultation, discussion, collaboration, webinar, learning materials, assessment of the knowledge, and communication in groups. Forms and types of interaction are closely related.

Teacher in a cloud-oriented sustainable learning environment in a higher education institution can inform students about a particular event, tools of adding news and calendar events; subgroups or groups of students. In particular, the supervisor can inform students of the problem group about the extraordinary meeting, etc.

Through consultations, students can get answers to questions that interest them, whether from a teacher in a particular subject or from a research supervisor for writing an article or coursework (diploma) project.

Another important form of interaction is the discussion, where students and the teacher (supervisor) are equal subjects of learning, resulting in the formation of their own opinion and the opportunity to defend it in a subgroup, group or team. This is also closely related to the form of group communication. Students can communicate through correspondence, chats.

In the process of performing laboratory work students often need the help of classmates, teachers. In this case, cooperation is a profitable solution. In cooperation, students develop such personal qualities as the ability to work in a team, sociability, etc.

To conduct online seminars for problem groups in such an environment is used webinar. This is quite a useful opportunity during the holidays and quarantine.

For successful mastering of the material, there is an opportunity to receive educational materials (lectures, theoretical information, literature, etc.). It is also a form of educational interaction, without which the learning process as a whole is not possible.

Evaluation of knowledge – is a form of interaction between actors, which is not possible without learning. That is why in a cloud-oriented sustainable learning environment of a higher education institution, this form of interaction is envisaged for further e-journaling and rating of a specific subject.

It should be noted that the main tools of communication are: dialogue, brainstorming, discussion, debate, debate, and their application transforms the educational process into mutual learning, where the student and the teacher are equal subjects of learning.

## 4 Conclusions

The designed cloud-oriented sustainable learning environment of a higher education institution should optimally address the challenges facing higher education institutions:

1. Planning of the educational process on different curricula and forms of study (full-time, part-time).
2. Organization of the educational process.
3. Organization of research work.
4. Submission of teaching materials.
5. Ensuring interaction between all participants of the educational process in the higher education institution.
6. Provision of information for teachers and students in various fields.
7. Ensuring the distribution of user access rights.
8. Organization of communities.
9. Ensuring that all necessary materials are shared.
10. Providing management of the educational process of preparation of future professionals.

Obviously, the main benefits of this environment for higher education institutions include: saving money on purchasing licensed software and not just software (everyone can use Office technology online); reducing the need for specially equipped premises; carrying out various types of educational work, control and assessment of knowledge online; saving computer memory (disk space); antivirus security of the educational environment; openness of the learning environment for teachers and students.

The design and application of a cloud-oriented sustainable learning environment with cloud LMS will provide such capabilities: keep electronic journals; use on-line services for the educational process; conduct correspondence, testing, and assessment of knowledge on-line; possibility of distance learning, library of books, manuals, textbooks, media files; file repositories; conference video and more. And all of the above is the key to a sustainable development of the learning environment.